\documentclass[reprint,aps,prd,superscriptaddress,preprintnumbers,nofootinbib]{revtex4-1}

\usepackage{amsfonts}
\usepackage{amsmath}
\usepackage{amsthm}
\usepackage{amssymb}
\usepackage{array}
\usepackage{dcolumn}
\usepackage{placeins}
\usepackage[svgnames]{xcolor}
\usepackage[hidelinks]{hyperref}
\hypersetup{
    colorlinks=true,
    linkcolor=NavyBlue,
    filecolor=NavyBlue,
    urlcolor=NavyBlue,
    citecolor=NavyBlue,
    }

\usepackage{graphics}
\usepackage{tikz}
\usepackage{graphicx}
\usetikzlibrary{positioning}
\usepackage{caption}
\usepackage{subcaption}
\usepackage{adjustbox}
\usepackage{gensymb}
\usepackage{breqn}
\usepackage{braket}
\usepackage{enumitem}
\usepackage{makecell}
\usepackage{ragged2e}
\usepackage{comment}
\usepackage{placeins}

\def\be{\begin{equation}}
\def\ee{\end{equation}}
\def\bea{\begin{eqnarray}}
\def\eea{\end{eqnarray}}

\def\tev{\, {\rm TeV}}
\def\gev{\, {\rm GeV}}
\def\mev{\, {\rm MeV}}

\newcommand{\gsim}{\lower.7ex\hbox{$\;\stackrel{\textstyle>}{\sim}\;$}}
\newcommand{\lsim}{\lower.7ex\hbox{$\;\stackrel{\textstyle<}{\sim}\;$}}

\begin{document}

\hfill \preprint{ MI-TH-219,  UH511-1319-2021}

\title{Explaining $g_{\mu}-2$ and $R_{K^{(*)}}$ using the light mediators of $U(1)_{T3R}$ }

\author{Bhaskar Dutta}
\email{dutta@physics.tamu.edu} 

\author{Sumit Ghosh}
\email{ghosh@tamu.edu}
\affiliation{Mitchell Institute for Fundamental Physics and Astronomy$,$ Department~ of ~Physics ~ and~ Astronomy$,$\\ Texas A$\&$M University$,$~College~ Station$,$ ~Texas ~77843$,$~ USA}

\author{Peisi Huang}
\email{peisi.huang@unl.edu}
\affiliation{Department of Physics  and Astronomy$,$\\  University of Nebraska$,$ Lincoln$,$ NE 68588$,$~ USA}

\author{Jason Kumar}
\email{jkumar@hawaii.edu}
\affiliation{Department of Physics and Astronomy$,$~ University~ of~ Hawaii$,$~ Honolulu$,$~ Hawaii~ 96822$,$~ USA}


\begin{abstract}
   Scenarios in which right-handed light Standard 
   Model fermions couple to a new gauge group, 
   $U(1)_{T3R}$ can naturally generate a sub-GeV 
   dark matter candidate.  But such models necessarily 
   have large couplings to the Standard Model, generally yielding tight experimental constraints.
   We show that the contributions to $g_\mu-2$ from the dark photon 
   and dark Higgs largely cancel out in the narrow window where all the experimental constraints are satisfied, leaving a net 
   correction which is consistent with recent 
   measurements from Fermilab.
   These models inherently violate lepton universality, and UV completions of these 
   models can include quark flavor violation which can explain $R_{K^{(\ast)}}$ anomalies as observed at the LHCb experiment after satisfying constraints on  $Br(B_s\rightarrow\mu^+\mu^-)$ and various other  constraints in the allowed parameter space of the model.   
 This scenario can 
   be probed by FASER, SeaQuest, SHiP, LHCb, Belle, etc.
\end{abstract}

\maketitle


\section{Introduction}

The $g_\mu-2$ anomaly has 
been one of the most 
promising signals of possible new physics beyond the 
Standard Model (SM)~\cite{Bennett:2006fi, Tanabashi:2018oca, Davier:2019can, Davier:2017zfy, Blum:2018mom, Keshavarzi:2018mgv, Blum:2019ugy, Campanario:2019mjh}.  
There are a variety of new physics 
scenarios which can potentially explain this anomaly, 
and which typically rely either on new heavy particles 
with a large coupling to muons, or light particles 
with a very small coupling to muons.  But there is 
an interesting scenario in which right-handed muons 
and other first- or second-generation fermions are 
charged under a new gauge group, 
$U(1)_{T3R}$~\cite{Dutta:2019fxn, Dutta:2020enk,Dutta:2020jsy}.  In this scenario, 
the symmetry-breaking scale of $U(1)_{T3R}$ 
($\sim {\cal O}(10~\gev)$) naturally feeds into the 
light SM fermion mass parameters, as well as the 
dark sector, yielding a sub-GeV dark matter 
candidate.  But the blessing is also a curse, as in 
this scenario the mediators  inherently have 
a large coupling to 
the SM, resulting in tight experimental constraints, 
and a typically very large correction to $g_\mu-2$.  There is only a small window in which the model is not 
ruled out by current laboratory, astrophysical, and 
cosmological observables.  But within this narrow 
window, there is a region of parameter space in 
which the dark Higgs ($\phi'$) 
and dark photon ($A'$) contributions 
to $g_\mu-2$ largely cancel, yielding a 
net contribution 
to $g_\mu-2$ which is consistent with the newest 
measurement from Fermilab. 

The combined data of Fermilab~\cite{Abi:2021gix} and BNL~\cite{Bennett:2006fi} increases the tension between the experimental value and the theoretical prediction~\cite{aoyama:2012wk, Aoyama:2019ryr, czarnecki:2002nt, gnendiger:2013pva, Davier:2017zfy, Blum:2018mom,  Keshavarzi:2018mgv, colangelo:2018mtw, hoferichter:2019gzf, Davier:2019can, Keshavarzi:2019abf, kurz:2014wya, melnikov:2003xd, masjuan:2017tvw, Colangelo:2017fiz, hoferichter:2018kwz, gerardin:2019vio, bijnens:2019ghy, colangelo:2019uex, Blum:2019ugy, colangelo:2014qya, Campanario:2019mjh} to 4.2 $\sigma$ level. This is given by, 

\bea \Delta a_\mu = a_\mu^{\text{exp}}-a_\mu^{\text{th}}=  (2.52 \pm 0.59) \times 10^{-9} \eea
The tension is less significant as claimed in a recent lattice calculation~\cite{Borsanyi:2020mff} which needs to be investigated further~\cite{Crivellin:2020zul,Keshavarzi:2020bfy,Colangelo:2020lcg}. 

The mass terms for fermions charged under $U(1)_{T3R}$ arise from 
non-renormalizable operators at the electroweak symmetry-breaking scale.  
A variety of UV completions of these models are possible, which generically 
permit quark flavor-violating processes involving heavy particles.  
On the other hand, since $U(1)_{T3R}$ couples to one complete generation, it 
necessarily induces lepton flavor non-universality through processes mediated 
by light mediators.  These processes together can generate anomalous lepton 
non-universality in $b$ decays, which can potentially explain the 
 recently observed $R_{K^{(\ast)}}$ anomalies~\cite{RKstar,RKstarBelle, Aaij:2021vac}. 
 These anomalies are very clean observables since they are devoid of  hadronic unceratinities.  
Very recently, using the full Run-1 and Run-2 data set, the  LHCb collaboration updated the $R_K$ result which now shows  a 3.1 $\sigma$ deviation from the SM~\cite{Aaij:2021vac}.  The full data analysis shows,
\begin{equation}
   R_K = 0.846 ^{+0.042}_{-0.039}\textrm{(stat)}^{+0.013}_{-0.012}\textrm{(syst)}, 
\label{rklhcb}\end{equation}
where the SM calculation yields $R_K=1.00\pm0.01$~\cite{Hiller:2003js,Bouchard:2013mia,Bordone:2016gaq}. 

Since the $g_{\mu}-2$ excess and the $R_{K^{(\ast)}}$ anomalies are indications of nonuniversality in the muon sector, it would be interesting to accommodate both of them in the context of a  model (for recent work, see ~\cite{Darme:2020hpo,Alvarado:2021nxy}). Such a model, however, can be  constrained by various other experimental data. For example, the CCFR constraint on $\nu_\mu N\rightarrow\nu_\mu N \mu^+\mu^-$ interaction~\cite{Mishra:1991bv} makes it difficult for a model to explain both anomalies~\cite{Bauer:2018onh,Altmannshofer:2019zhy}.  Measurement of  $Br(B\rightarrow K^{(*)}\nu\nu)$ also  restrict the parameter space. Both these neutrino related measurements constrain models which utilize left-handed muons to solve the $R_{K^{(\ast)}}$ puzzles. Additionally, the  measurements of $Br(B_s\rightarrow\mu^+\mu^-)$  restricts the parameter space of such models. 
In the context of the $U(1)_{T3R}$ model, where the new gauge boson  does not couple to the left-handed neutrino, we will show how both anomalies can be accommodated  after satisfying various  experimental data including the recent muon $g-2$ result. We also show the predictions for a few more $B$ decay observables  which can test this model with more data, new measurements and improved theoretical understanding of form factors. 

A good way to probe the allowed parameter space of this scenario experimentally is 
at beam dump experiments with a displaced detector, 
where one can search for the decays of the long-lived 
dark photon to $e^+ e^-$.  The difficulty is that, 
because these models have a relatively large coupling 
to the Standard Models, the decay length of the 
$A'$ tends to be shorter than typically expected; 
although it exits the immediate interaction region, 
it often will decay before reaching many displaced 
detectors.  Thus, there are regions of parameter space
($m_{\phi'} \sim 70-90\mev$, $m_{A'} \sim 60-200\mev$)
in which the measured 
value of $g_\mu-2$ can be explained, and  
which lie just beyond current bounds from U70/NuCal~\cite{Bauer:2018onh, Davier:1989wz, Gninenko:2014pea}.  
Portions of this region can be probed by 
FASER~\cite{Feng:2017uoz, Ariga:2018zuc, Ariga:2018pin, Ariga:2018uku, Ariga:2019ufm}, SeaQuest~\cite{Berlin:2018pwi, Aidala:2017ofy}, and SHiP~\cite{Anelli:2015pba, Alekhin:2015byh}. 
There is a portion of 
this parameter space ($m_{A'} > 110\mev$) which 
cannot be explored by even these experiments.

Alternatively, if the $A'$ decays invisibly, then there is a region of parameter space 
($m_{\phi'} \sim 95-102\mev$, $m_{A'} \sim 10-30\mev$) in which scenario will again 
evade all constraints from laboratory experiments and cosmological observables, while 
also yield a prediction for $g_\mu-2$ which is consistent with experiment.  
Interestingly, this scenario can also potentially explain an excess event rate seen by 
the COHERENT experiment, and this mass range can be probed by the upcoming 
NA-64$\mu$ and LDMX-M${}^3$ experiments.

The plan of this paper is as follows.  In 
Section~\ref{sec:Model}, 
we review the $U(1)_{T3R}$ model and the contribution 
to $g_\mu-2$.  
In Section~\ref{sec:Allowed}, we discuss constraints 
on this scenario, and identify the regions of 
parameter space which are allowed, and the experiments 
which may further constrain this scenario.
In Section~\ref{sec:Flavor}, we discuss the explanations of the $R_{K^{(\ast)}}$ anomalies in the  allowed parameter space of the model and list various predictions.
In Section~\ref{sec:Conclusion}, we conclude.


\section{The Model and $g_\mu -2$}
\label{sec:Model}

We consider the scenario in which the right-handed 
$\mu$, $\nu$, $u$ and $d$ are charged under $U(1)_{T3R}$, with up-type and down-type fermions 
having opposite sign charges ($\pm 2$).  
In this scenario, all gauge anomalies automatically cancel.  Note, it is 
technically natural~\cite{Batell:2017kty} for the charged lepton and either the up-type or down-type 
quark charged under $U(1)_{T3R}$ to be mass eigenstates.  
The details of this 
model are described in~\cite{Dutta:2019fxn, Dutta:2020enk,Dutta:2020jsy}, and are reproduced 
in the Appendix.  

$U(1)_{T3R}$ is 
spontaneously broken to a parity  
by the condensation of the scalar field $\phi$.  
We denote by $V = 
\langle \phi \rangle$ the vacuum expectation 
value of $\phi$, while the dark Higgs $\phi'$ 
is the real field 
which denotes an excitation away from the vev.  
The dark photon $A'$ then gets a mass 
given by $m_{A'}^2 = 2 g_{T3R}^2 V^2$, where 
$g_{T3R}$ is the coupling of $U(1)_{T3R}$.  

Because 
the right-handed muon is charged under $U(1)_{T3R}$, 
while the left-handed muon is not, the muon mass is 
protected by $U(1)_{T3R}$.   As a result, in the low-energy effective field theory 
(below the scale of electroweak symmetry-breaking), $\phi$ couples 
to the muon as $\lambda_\mu 
\phi \bar \mu \mu$.  If we define  $\phi = V + (1/\sqrt{2}) \phi'$, 
then we find $m_\mu = \lambda_\mu V$; if we choose $V = {\cal O}(10\gev)$, 
then the effective Yukawa coupling $\lambda_\mu$ is not unnaturally small.
The dark Higgs then couples to muons with a coupling 
$m_\mu / \sqrt{2} V$.

The dark sector also includes a Dirac fermion $\eta$ with charge $1$ 
under $U(1)_{T3R}$.  This fermion can get a Majorana mass term through a 
coupling to $\phi$; if this is larger than the Dirac mass, then 
one is left with two dark sector Majorana fermions, $\eta_{1,2}$ with 
masses proportional to $V$ and a small mass splitting. These fermions also couple to $\phi'$ and $A'$.  The
$\eta_{1,2}$ are the 
only particles which are odd under the surviving parity, and the lightest 
of these is a dark matter candidate with a mass which is 
naturally sub-GeV.

The new fields added in this model are $A'$, $\phi'$, 
$\eta_{1,2}$, and $\nu_R$.  We assume that the 
sterile neutrino $\nu_S$ 
is mostly $\nu_R$, with only a 
very small mixing with left-handed neutrinos.  If the 
sterile neutrino is reasonably light, it can be 
relevant in formulating constraints on this scenario.  
For simplicity, we will assume that it is moderately 
heavy, and plays little role in these constraints.

We thus see that, once we specify $V$, $m_{A'}$ and 
$m_{\phi'}$, the coupling of the dark photon and 
the dark Higgs to the muon are fixed.  We will set 
$V = 10\gev$, following~\cite{Dutta:2019fxn}, and 
consider the correction to the muon magnetic moment 
as a function of $m_{A'}$ and $m_{\phi'}$.

\begin{figure}[h]
\centering
\includegraphics[height=5cm,width=5cm]{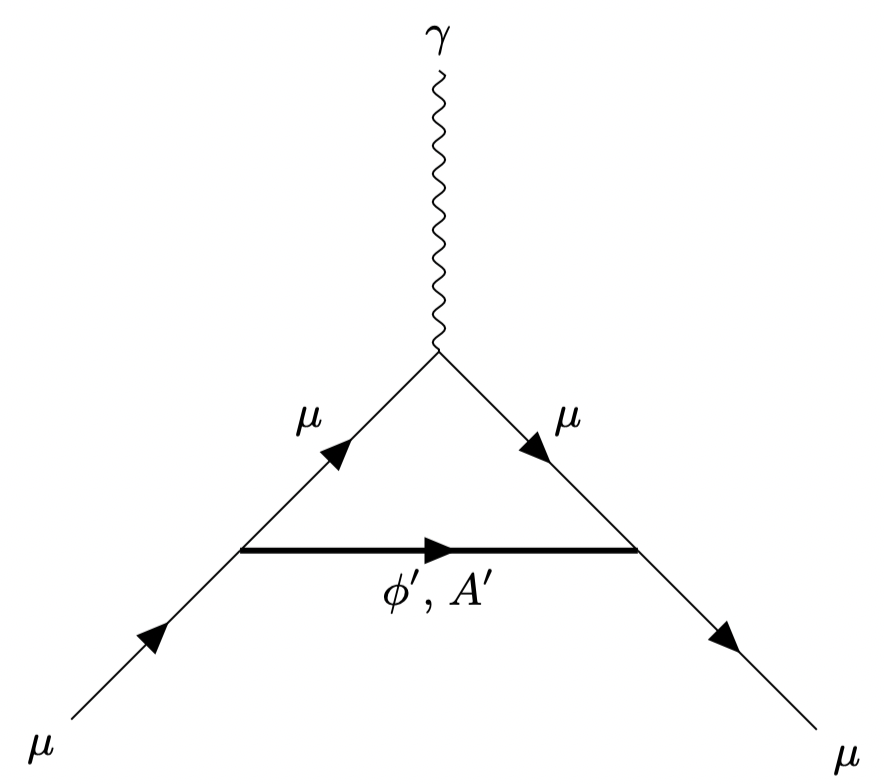}
\captionsetup{justification   = RaggedRight,
             labelfont = bf}
\caption{\label{fig:g2T3Rscalar} One-loop $\phi^\prime$/$A^\prime$ contribution to $g_\mu$-2. }
\label{fig:phi'}
\end{figure}

\begin{figure}[h]
\centering
\includegraphics[height=5.5cm,width=7.5cm]{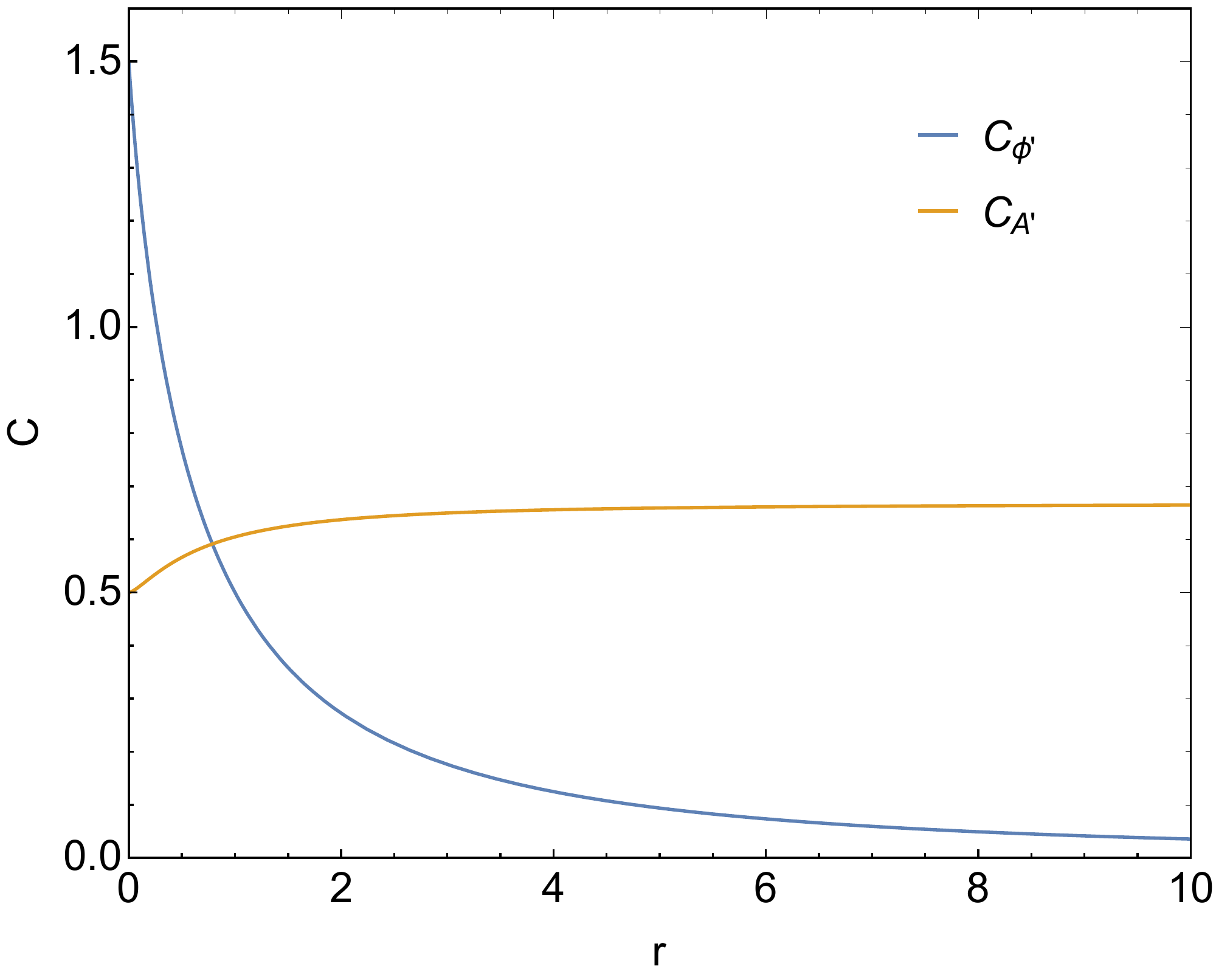}
\captionsetup{justification   = RaggedRight,
             labelfont = bf}
\caption{\label{fig:Cvsrplot}  
Plot of $C_{\phi'}$ and $C_{A'}$ as functions of $r_{\phi'}$ and 
$r_{A'}$, respectively.}
\end{figure}

The muon anomalous magnetic moment will 
receive corrections arising from diagrams in which 
either $\phi'$ or $A'$ run in the loop (see Figure~\ref{fig:phi'}). 
The correction to $a_\mu = (g_\mu -2)/2$  due to one-loop diagrams involving $A'$ and $\phi'$ is given by~\cite{Leveille:1977rc} 
\bea 
\label{Delta} 
\Delta a_\mu &=& \frac{m_\mu^4}{16\pi^2V^2}  \int_0^1 dx \frac{(1-x)^2(1+x)}{(1-x)^2m_\mu^2+xm_{\phi^\prime}^2}   
\nonumber\\  
&\,& + \frac{m_\mu^2 }{16\pi^2V^2} \int_0^1 dx \frac{x(1-x)(x-2)m_{A^\prime}^2-x^3m_\mu^2}{x^2m_\mu^2+(1-x)m_{A^\prime}^2} ,
\nonumber\\
&=& (6.98 \times 10^{-7}) \left( \frac{V}{10\gev} \right)^{-2} \left(C_{\phi'} - C_{A'} \right) ,
\eea
where
\bea
C_{\phi'} &=&  
\int_0^1 dx \frac{(1-x)^2(1+x)}{(1-x)^2+x r_{\phi'}^2} 
\nonumber\\
C_{A'} &=&   \int_0^1 dx \frac{x(1-x)(2-x) r_{A'}^2 +x^3}{x^2+(1-x)r_{A^\prime}^2} ,
\eea
and $r_{\phi'} \equiv m_{\phi'}/m_\mu$, 
$r_{A'} \equiv m_{A'}/m_\mu$.  The contribution to $g_\mu-2$ from $\phi'$ is always positive, 
while the contribution from $A'$ is always negative, as the $A'$ has both vector and axial 
couplings to the muon.  These contributions must cancel to within ${\cal O}(1\%)$ in order for 
the total correction to $g_\mu-2$ to be consistent with observations. We plot 
$C_{\phi'}(r_{\phi'})$ and $C_{A'}(r_{A'})$ in Figure~\ref{fig:Cvsrplot}.

Interestingly, the contribution from the $A'$ is nearly universal; $C_{A'}$ only varies 
between $1/2$ and $2/3$.  In particular, as $m_{A'}$ grows, the coupling also grows and the 
contribution to $g_\mu -2$ asymptotes to a constant.  But even as $m_{A'}$ decreases and the gauge
coupling goes to zero, the contribution to $g_\mu-2$ still asymptotes to a constant, because 
the longitudinal polarization effectively becomes a pseudoscalar Goldstone mode, with the same 
coupling to muons as the $\phi'$.  As such, $g_\mu-2$ can only be consistent with experiment 
for $C_{\phi'}$ between $1/2$ and $2/3$, which corresponds $(2/3) m_\mu \lesssim 
m_{\phi'} \lesssim m_\mu$.
We thus see that one can 
only obtain consistency with $g_\mu-2$ measurements 
for $m_{\phi'}$ within the very narrow range of 
$\sim 67 - 100\mev$.  Note that, two-loop Barr-Zee type diagrams are also possible with dark Higgs/photon in this scenario. But the contribution is negligible compared to the one-loop contribution~\cite{Harnik:2012pb, Ilisie:2015tra}.

Thus far, we have only considered the low-energy effective 
field theory defined below the scale of electroweak symmetry 
breaking.  
But the muon mass term requires both a 
Higgs and a dark Higgs insertion, and thus 
arises from a non-renormalizable operator of 
the form $(1/\Lambda)H \phi \bar \mu \mu$ in 
the theory defined above electroweak symmetry-breaking.
There should be some UV-completion of this theory, 
and one might wonder if the corrections to 
$g_\mu-2$ induced by the new UV fields could spoil 
the result we have found.  To explore this, we 
consider, as an example of a possible UV-completion, 
the universal seesaw~\cite{Berezhiani:1983hm,Chang:1986bp,Davidson:1987mh,DePace:1987iu,Rajpoot:1987fca,Babu:1988mw,Babu:1989rb,Babu:2018vrl}.  
In this case, there is a new 
heavy fermion $\chi_\mu$, charged under hypercharge, which couples to muons as 
$\lambda_L H \bar \chi_\mu P_L \mu +  
\lambda_R \phi \bar \chi_\mu P_R \mu +h.c.$.  In the 
theory defined below $m_{\chi_\mu}$, this will yield 
the required effective operator, subject to the 
seesaw relation $m_\mu \sim \lambda_L \lambda_R 
\langle H \rangle V / m_{\chi_\mu}$.  If we 
take $V=10\gev$ and $\lambda_{L,R} = {\cal O}(1)$, 
then we find $m_{\chi_\mu} = {\cal O}(10\tev)$; the 
corrections yielded by introducing this field will 
not substantially change our discussion.  
Note that if $\lambda_L \lambda_R$ is significantly 
smaller than unity, then $\chi_\mu$ may be 
light enough to be probed at the LHC.


\section{Allowed Regions of Parameter Space and 
Future Probes}
\label{sec:Allowed}

We will now consider the regions of this parameter space are 
consistent with other laboratory experiments.  
The relevant experiments are 
those in which the $A'$ and $\phi'$ are produced at accelerator  
experiments and either decay invisibly, or decay visibly at 
displaced detectors~\cite{Dutta:2020enk}.  

If the $A'$ decays invisibly, then this 
scenario would be ruled out by data from COHERENT~\cite{Akimov:2017ade, Akimov:2018vzs, Akimov:2018ghi, Akimov:2019xdj, Akimov:2020pdx} and Crystal Barrel~\cite{Amsler:1994gt, Amsler:1996hb}, 
unless $m_{A'} < 30\mev$.  
But for $m_{A'} \lesssim {\cal O}(10\mev)$, this scenario faces tension with 
cosmological bounds (corrections to $N_{eff}$~\cite{Aghanim:2018eyx, Dutta:2020jsy}), although 
there are more complicated scenarios in which this tension can be alleviated.  For 
$m_{A'}$ in the $\sim 10 - 30\mev$ range, $A'$ production 
can potentially contribute to anomalous supernova cooling
~\cite{Harnik:2012ni, Redondo:2008aa, Bollig:2020xdr, Croon:2020lrf}).  But in this mass range, the 
coupling $g_{T3R}$ is large enough that the $A'$ will decay promptly, and the decay 
products will not be able to free-stream out of the supernova.  If $A'$ in the $\sim 10 - 30\mev$ mass 
range decays invisibly, it will satisfy all other current laboratory and cosmological constraints, and the constraints 
on $g_\mu -2$ will also be satisfied for 
$m_{\phi'}$ in the $\sim 95 - 102 \mev$ range.  

Moreover, for $m_{A'} \sim 30\mev$, this scenario can explain the excess of events seen by the 
COHERENT experiment.  The COHERENT experiment collides a 
proton beam against a fixed target, and searches for the 
scattering of neutrinos produced by these collisions at 
a displaced detector.  The COHERENT experiment sees a 
$2.4-3\sigma$ excess of events~\cite{Dutta:2019nbn}, 
which could be explained 
if the $A'$ ($m_{A'} \sim 30\mev$) decays to either 
dark matter or sterile neutrinos, which in turn scatter 
against nuclei in the distant detector by $A'$ exchange~\cite{Dutta:2020enk}.  

This scenario, in which the 
$A'$ ($m_{A'} \in [10,30]\mev$) and 
$\phi'$ ($m_{\phi'} \in [95,102]\mev$) decay invisibly, 
can be probed definitively by the upcoming 
NA-64$\mu$ and LDMX-M${}^3$ experiments.
We will address the scenario of invisible $A'$ decay 
further, in the context of flavor anomalies.

\begin{figure}[h]
\centering
\includegraphics[height=6cm,width=8cm]{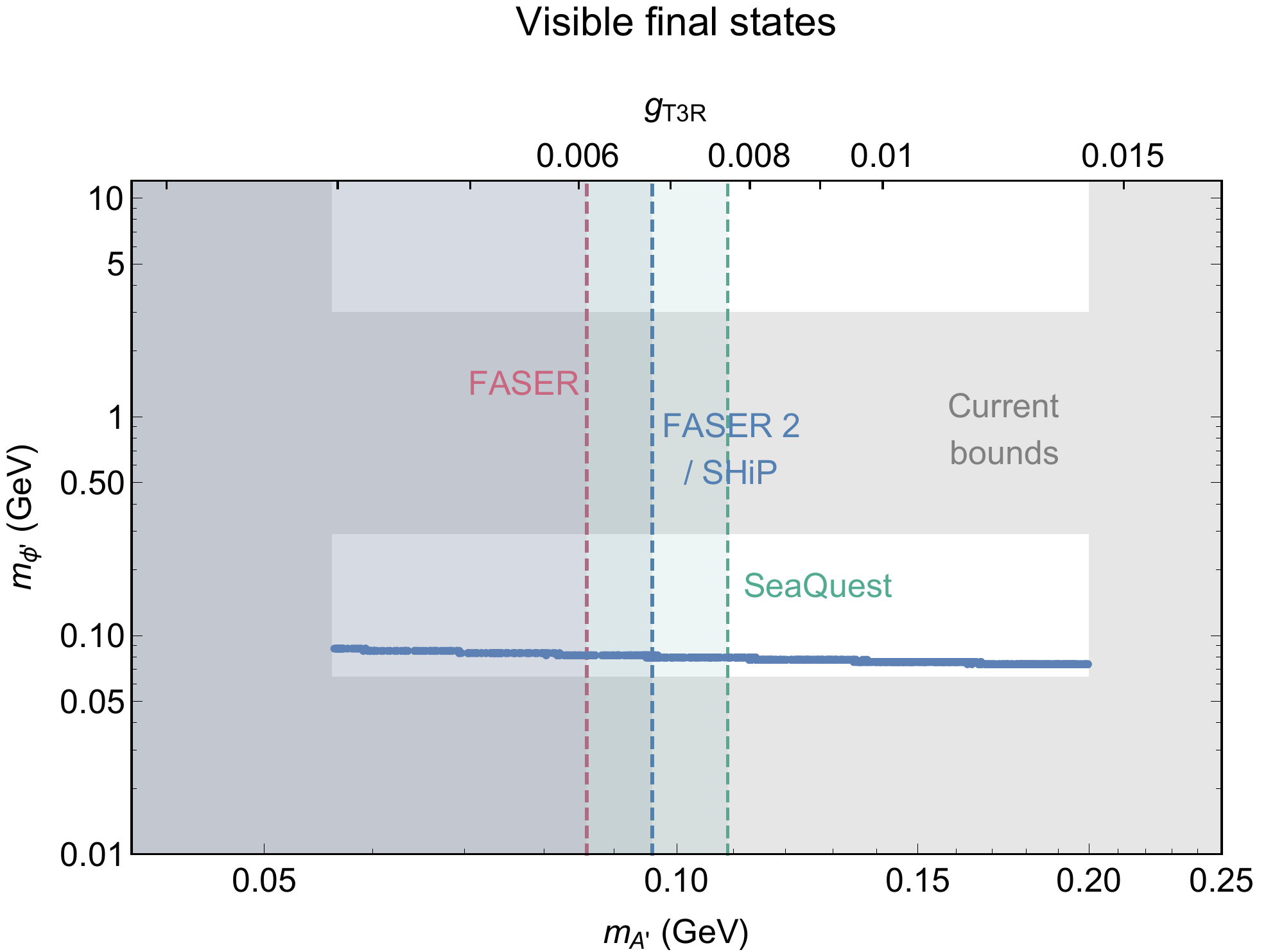}
\captionsetup{justification   = RaggedRight,
             labelfont = bf}
\caption{\label{fig:VisibleSensitivity} Plot of the region in 
the $(m_{A'}, m_{\phi'})$-plane which is consistent with current 
measurments of $g_\mu-2$ (blue), along with current exclusion bounds 
(grey) from U70/NuCal, E137, and Babar, and the future sensivity of 
FASER (red transparent), FASER 2/SHiP (blue transparent) and SeaQuest (green 
transparent).  $g_{T3R}$ is shown on the top axis.}
\end{figure}

The visible decay $A' \rightarrow e^+ e^-$ can be mediated by 
one-loop kinetic mixing, in which the right-handed SM fermions charged 
under $U(1)_{T3R}$ run in the loop.  Assuming no tree-level kinetic mixing, we find a $\gamma -A'$ kinetic mixing parameter of 
$\epsilon \sim (m_{A'} / \sqrt{2}V) 
\sqrt{\alpha_{em} / 4\pi^3}$.
This scenario is ruled out by data from U70/NuCal unless $m_{A'} > 56\mev$, and 
by data from BaBar~\cite{Aubert:2009cp, Lees:2014xha, Bauer:2018onh} unless $m_{A'} < 200\mev$.
For $m_{A'}$ in the $56-200\mev$ range, 
the range of $m_{\phi'}$ for which $g_\mu-2$ can match observation 
is indeed very narrow: $74-86 \mev$.  
We plot the region of $(m_{\phi'}, m_{A'})$ parameter space consistent 
with current $g_\mu-2$ observations in Figure~\ref{fig:VisibleSensitivity}, along 
with current bounds from U70/NuCal, BaBar and E137~\cite{Riordan:1987aw, Bjorken:1988as, Bjorken:2009mm}.  
We also plot the sensitivity of 
FASER, FASER 2, SHiP and SeaQuest, which also can 
search for the displaced decays of $A'$.  These 
bounds and sensitivities are discussed in more 
detail in~\cite{Dutta:2020enk}.

Note that, if we increase the value of $V$, then 
we will reduce the precision with which $C_{\phi'}$ 
and $C_{A'}$ must cancel.  But increasing $V$ will 
also result in a longer lifetime for $A'$, since it 
would yield a reduced gauge coupling for $U(1)_{T3R}$. This would raise the  lower bound on $m_{A'}$ from U70/NuCal, which is determined by fact that, for larger $m_{A'}$, the dark photon decays before it reaches 
the detector.  Thus, one cannot significantly  
reduce the 
precision of the required cancellation by increasing 
$V$.

\subsection{Dark Matter 
Relic Density and Direct Detection}  
\label{sec:Relic}

The motivation for coupling the light 
fermions to $U(1)_{T3R}$ was for the new 
light scale to not only feed into the light 
SM fermion masses, but also the dark sector, 
providing a predictive framework for determining 
the dark matter mass scale.  
The dark sector consists of two Majorana 
fermions, $\eta_{1,2}$ which couple to 
$\phi'$ ($\propto m_\eta / V$) and 
$A'$ ($\propto m_{A'}/V$)  These couplings 
allow the dark particles to interact with 
the SM, potentially diluting the relic density, 
and yielding a direct detection signal.

Dark matter co-annihilation in the early Universe 
can be mediated by the $A'$, but the only accessbile 
final states are $\nu_A \nu_A$ and $e^+ e^-$ (the 
$\gamma \gamma$ final state is forbidden by the 
Landau-Yang Theorem~\cite{Yang:1950rg}).  Since both of these final 
states are suppressed, either by a neutrino mixing 
angle or a kinetic mixing parameter, co-annihilation 
via an intermediate $A'$ will play no role in our 
benchmark scenario.

For $m_\eta \sim 
(1/2) m_{\phi'}$, the relic density can instead be 
sufficiently depleted by annihilation to photons 
via the $\phi'$ resonance ($\eta \eta \rightarrow 
\phi' \rightarrow \gamma \gamma$) to match current 
observations.  Note that this annihilation 
cross section is $p$-wave suppressed.  This 
suppression was only an ${\cal O}(10)$ factor at 
the time of dark matter freeze-out, but was much 
larger at late times, leading to a negligible 
contribution to CMB distortions or current 
indirect detection signatures.

Dark matter spin-independent velocity-independent 
scattering with 
nuclei can be mediated by either the $\phi'$ or 
$A'$, which in this scenario couple both to the 
dark matter and to $u$-/$d$-quarks.  Scattering 
mediated by $\phi'$ is elastic and isospin-invariant, 
while scattering mediated by the $A'$ is 
inelastic (since Majorana fermions can only have off-diagonal vector couplings) and maximally 
isospin-violating~\cite{Chang:2010yk,Feng:2011vu,Feng:2013vod} (since 
the $A'$ couples to $u$ and $d$ with opposite signs).

\begin{figure}[h]
\centering
\includegraphics[height=5.7cm,width=7.7cm]{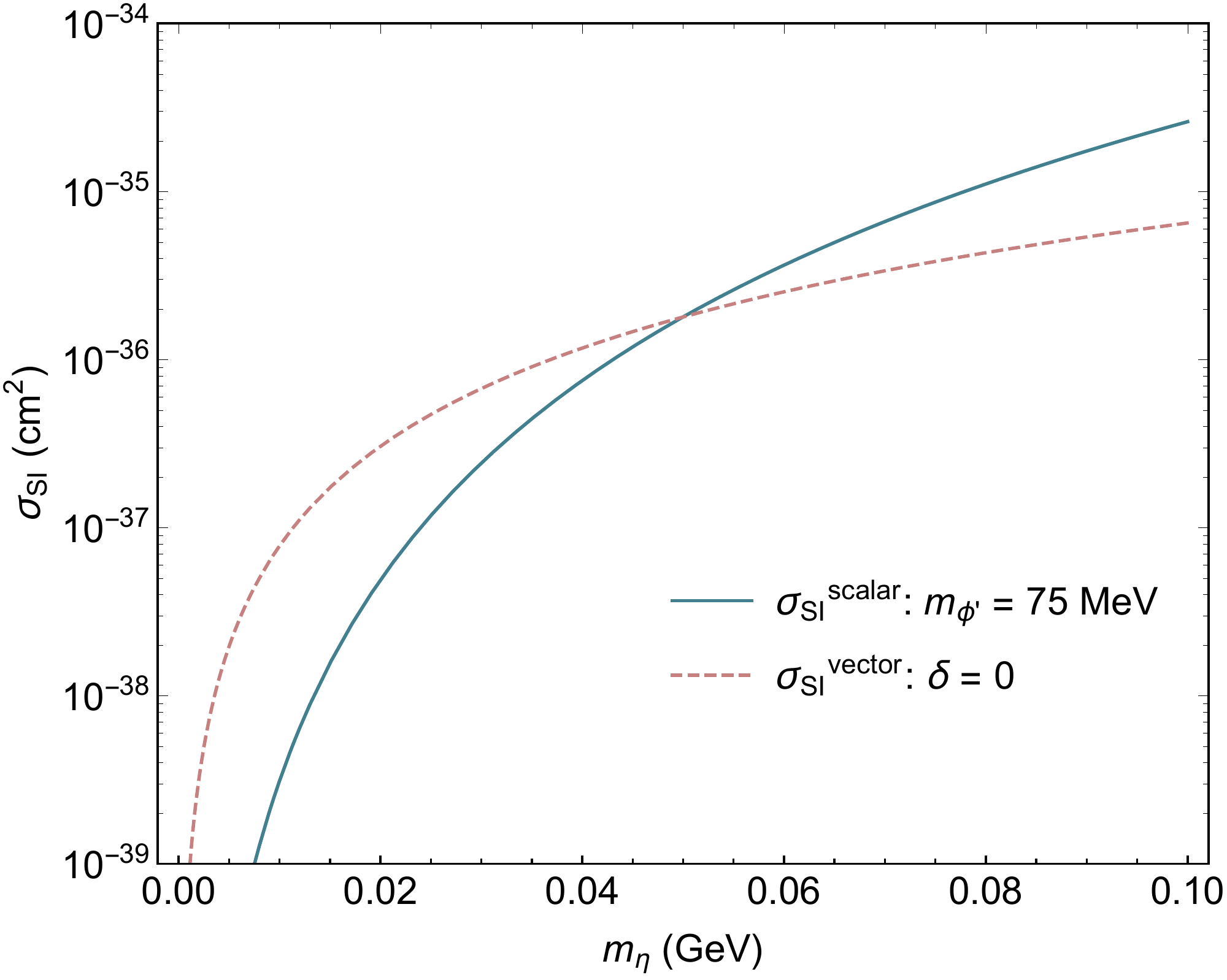}
\captionsetup{justification   = RaggedRight,
             labelfont = bf}
\caption{\label{fig:ddplot}  
Spin-independrnt dark matter-nucleon scattering cross section as a function of the dark matter mass. The blue (solid) line shows the SI elastic scattering cross section mediated by $\phi^\prime$ for $m_{\phi^\prime}=$ 75 MeV. And the orange (dashed) line shows the SI inelastic scattering cross section mediated by $A^\prime$ with $\delta=0$. }
\end{figure}

In Figure~\ref{fig:ddplot}, we plot the 
elastic spin-independent dark matter nucleon 
scattering cross section mediated by $\phi'$, 
assuming $m_{\phi'} = 75\mev$.  In the same plot, we also show the inelastic spin-dependent dark 
matter proton scattering cross section mediated 
by $A'$, assuming that the dark matter mass 
splitting is negligible.  
Note that this cross section is independent of 
$m_{A'}$ at fixed $V$, since $g_{T3R} \propto m_{A'}$.
For $m_{\eta_1} \lesssim 100\mev$, these scenarios 
are unconstrained by current direct detection 
experiments~\cite{Abdelhameed:2019hmk, Bringmann:2018cvk, Dent:2019krz, Liu:2019kzq}.

\section{Flavor Anomalies}
\label{sec:Flavor}

This model  
can  potentially impact the variety of 
anomalies in observables based on the process $b \rightarrow s 
\ell^+ \ell^-$.  If these anomalies are explained by new 
physics, it points to a scenario which generates both lepton 
flavor non-universality and quark flavor violation.  The scenario we 
describe here can contribute to both of the above, implying that 
it may also contribute to these flavor anomalies.

Lepton flavor non-universality arises from the low-energy sector of the 
theory, since the $\phi'$ and $A'$ couple only to $\mu$ at tree-level.  
On the other hand, quark flavor violation 
can arise from the UV completion of this model.
We have considered, as possible UV completions, 
the addition of heavy fermions which have 
same EM charge as SM fermions, but different 
charges under $SU(2)_L$ and/or 
$U(1)_{T3R}$. Once these heavy fermions are integrated
out, we generate the low energy effective Lagrangian as described in the Appendix A.

When these new fermions are added, 
the $Z$ and $A'$ couplings to fermions in 
the flavor eigenstate basis are diagonal 
matrices which need not be proportional to 
the identity.  As a result, these coupling 
matrices can become non-diagonal in the 
mass eigenstate basis, yielding vertices 
of the form $\bar b \gamma^\mu 
P_{L,R} s (Z,A')_\mu$.  Note, such 
flavor-changing is not allowed for the photon 
coupling, as a result of gauge-invariance 
(in particular, the photon coupling matrix is proportional to the identity in every basis).
Terms of the form 
$\bar b \gamma^\mu P_{L,R} s Z_\mu$ can 
contribute to universal quark 
flavor-changing processes ($b \rightarrow s \ell^+ \ell^-$), while terms of the form 
$\bar b \gamma^\mu P_{L,R} s A'_\mu$ can 
contribute to lepton non-universal quark 
flavor-changing processes 
($b \rightarrow s \mu^+ \mu^-$).

\begin{figure}[h]
\begin{subfigure}[h]{0.3\textwidth}
\includegraphics[width=0.8\linewidth,height=5cm]{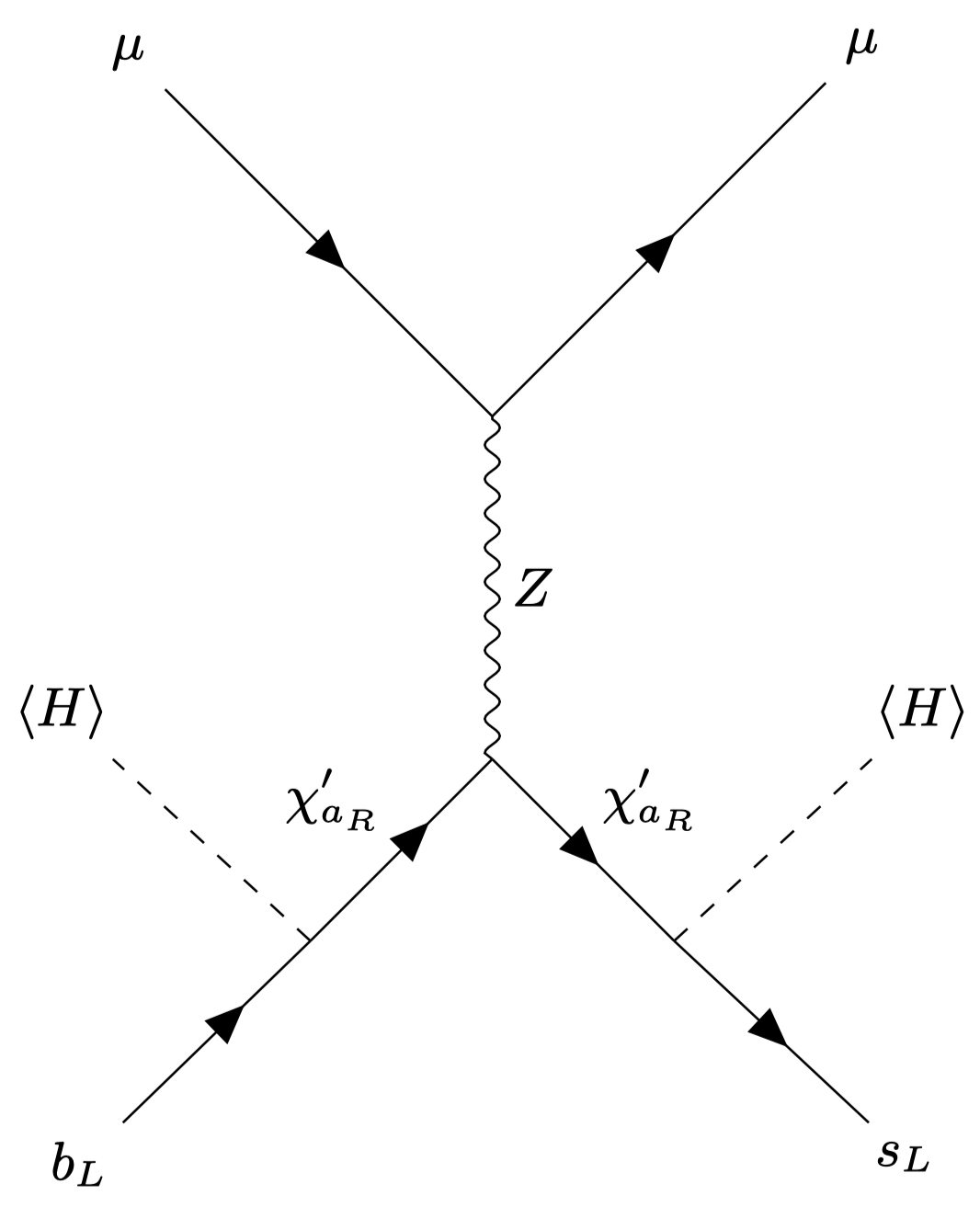}
\captionsetup{labelfont = bf}
\caption{\label{fig:RKU}}
\end{subfigure}	
\hspace{0.0cm}	
\begin{subfigure}[h]{0.3\textwidth}
\includegraphics[width=0.8\linewidth,height=5cm]{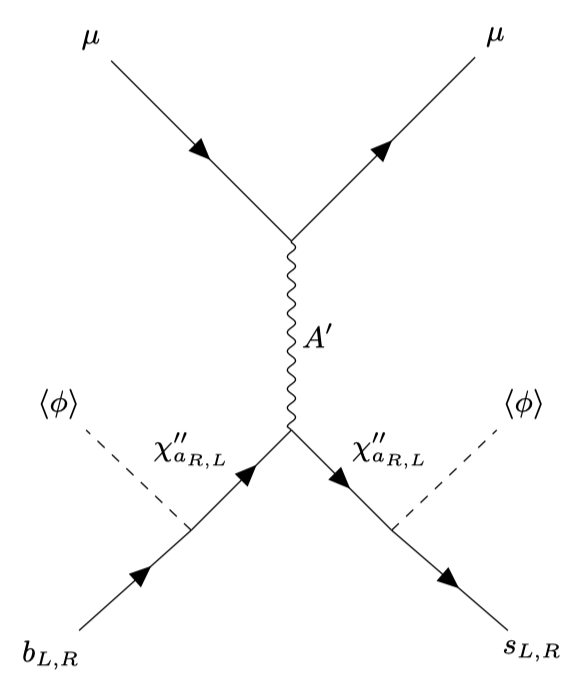}	
\captionsetup{labelfont = bf}
\caption{\label{fig:RKNU}}
\end{subfigure}	
\hspace{0.0cm}
\begin{subfigure}[h]{0.3\textwidth}
\includegraphics[width=0.8\linewidth,height=5cm]{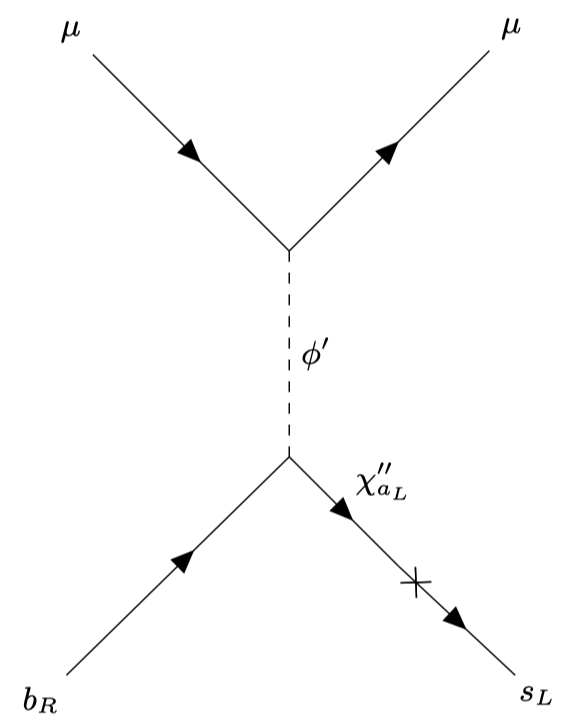}
\captionsetup{labelfont = bf}
\caption{\label{fig:RKscalar}}
\end{subfigure}
\captionsetup{justification   = RaggedRight,
             labelfont = bf}
\caption{\label{fig:RK} Feynman diagrams that contribute to the B-anomalies. } 
\end{figure}

As an example, 
we have considered a UV completion based on the universal seesaw~, in which one 
introduces new heavy vector-like fermions $\chi_{u,d,\mu,\nu}$ which are neutral under  $U(1)_{T3R}$, but have the same SM 
quantum numbers as $u_R$, $d_R$, $\mu_R$ and 
$\nu_R$, respectively.  While this is a minimal UV completion, one could add additional 
generations of these heavy particles, or even a single additional particle, without 
generating anomalies.  
Consider adding an additional $\chi'_a$, 
which mixes with $b$ and $s$ through Lagrangian 
terms 
of form $\lambda'_{b,s} H \bar Q_L^{b,s} P_R 
\chi'_a 
+ m'_{b,s}  \bar\chi'_a  P_R q_R^{b,s} 
+ h.c.$ (we assume negligible mixing with 
the first generation).  We see that  
$(\chi'_a)_R$ has same $Z$ coupling as 
$(b,s)_R$, while $(\chi'_a)_L$ has a $Z$ coupling which differs from $(b,s)_L$, and 
$(b,s,\chi'_a)_{L,R}$ are all neutral under 
$U(1)_{T3R}$.  In this scenario, we would 
find a vertex of the form 
$\bar b \gamma^\mu P_L s Z_\mu$ at tree-level 
(Fig.~\ref{fig:RKU}), 
but with no similar coupling for right-handed 
quarks (since the $Z$-coupling to the 
right-handed 
quarks is the identity in every basis).  
A coupling of the form 
$\bar b \gamma^\mu P_L s A'_\mu$ is also induced 
at one-loop through $Z-A'$ kinetic mixing, 
but this term will generally be small if 
the kinetic mixing is small.

\onecolumngrid

\twocolumngrid

By the same token, one could instead add a 
vector-like fermion $\chi''_a$  with the 
same SM quantum  numbers as $(b,s)_R$ 
but with $U(1)_{T3R}$ 
charge $Q_{T3R} = 2$.  This fermion could 
mix with $b,s$ (we assuming negligible mixing 
with $d$) through a Lagrangian term of the 
form $\lambda''_{b,s} \phi \bar \chi''_a 
P_R q_{b,s}$.  Since $\chi''_a$ is 
charged under $U(1)_{T3R}$ while $b,s$ are not, 
this term will yield a tree-level contribution 
to the coupling 
$\bar b \gamma^\mu P_{L,R} s A'_\mu$ (Fig.~\ref{fig:RKNU}).  Similarly, 
since $(\chi''_a)_L$ has a different $Z$ coupling 
than $(b,s)_L$, this term will yield a tree-level 
contribution to the coupling 
$\bar b \gamma^\mu P_L s Z_\mu$.  In this case, 
there is no similar contribution to 
$\bar b \gamma^\mu P_R s Z_\mu$, since 
$(b,s,\chi''_a)_R$ all have identical coupling 
to the $Z$ boson.  These considerations would 
be reversed if we had instead given the 
$\chi''_a$ the same SM gauge charges as 
$(b,s)_L$.

Note that the introduction of $\chi''_a$ will 
also induce a vertex of the form 
$\lambda''_{b,s} \phi' \bar q_{L(s,b)} q_{R(b,s)} 
\sin \theta'_{(s,b)L}$.  The related diagram is shown in Fig.~\ref{fig:RKscalar}.

We can approximate the effect of these 
interactions with effective operators 
which couple a ($b,s$) quark bilinear to a 
muon bilinear.  For diagrams in which 
$\phi'$ or $A'$ is exchanged, since the 
energy transfer is much larger than the 
mediator, we may approximate the energy scale 
of the operator with energy scale of the 
process, $\Lambda \sim {\cal O}(2\gev)$.

The diagrams which involve $\phi'$ 
exchange will contribute to effective operators 
with scalar Lorentz structure.  The diagrams 
which involve $Z$ or $A'$ exchange will 
contribute to effective 
operators with vector or axial-vector 
Lorentz structure, and also to operators with 
pseudoscalar structure (arising from the 
Goldstone mode, or equivalently, the chiral 
coupling of the longitudinal polarization).  
We may ignore this operator for the case of 
$Z$-exchange, however, since the mass of the 
gauge boson is much larger than the energy of 
the process.

The effective operator corresponding to Fig.~\ref{fig:RKU} can be written as,

\bea
{\cal O}_{U}^{Z} &=& 
\frac{e^2 }{3m_Z^2} 
\tan^2 \theta_W 
\left( \sin \theta_{sL} \sin \theta_{bL} + \sin \theta'_{sL} 
\sin \theta'_{bL} \right) 
\nonumber\\
&\,& 
\left(\bar b \gamma^\mu P_L s \right) 
\nonumber\\
&\,& 
\left(\bar \mu \gamma_\mu 
\left[P_R + \left( 1-
\frac{1}{2 \sin^2 \theta_W} \right) P_L \right] 
\mu \right)  \eea

We can express the effective operators corresponding  to Fig.~\ref{fig:RKNU} and \ref{fig:RKscalar} as
as,
\bea
{\cal O}^{A'}_{NU} &=& \frac{1}{\Lambda^2} 
\sin \theta'_{s(L,R)} \sin \theta'_{b(L,R)} \left(\frac{m_{A'}}{\sqrt{2} V} \right)^2 
\nonumber\\
&\,& 
\left(\bar b \gamma^\mu P_{L,R} s \right)(\bar \mu \gamma_\mu P_R \mu) 
\nonumber\\
&\,& +
\frac{1}{\Lambda^2} 
\sin \theta'_{s(R)} \sin \theta'_{b(R)} \left(\frac{m_\mu m_b}{2 V^2} \right) 
\nonumber\\
&\,& 
\left(\bar b \gamma^5 s \right)(\bar \mu \gamma^5 \mu) 
\nonumber\\
&\,& -
\frac{1}{\Lambda^2} 
\sin \theta'_{s(L)} \sin \theta'_{b(L)} \left(\frac{m_\mu m_s}{2 V^2} \right) 
\nonumber\\
&\,& 
\left(\bar b \gamma^5 s \right)(\bar \mu \gamma^5 \mu) , 
\eea

\bea {\cal O}^{\phi'}_{NU} &=& 
\frac{\lambda''_s}{\Lambda^2} 
\sin \theta'_{bL} \frac{m_\mu}{\sqrt{2} V}
(\bar b P_R s) (\bar \mu \mu) 
\nonumber\\
&\,& + \frac{\lambda''_b}{\Lambda^2} 
\sin \theta'_{sL} \frac{m_\mu}{\sqrt{2} V}
(\bar b P_L s) (\bar \mu \mu) ,
\eea
where $\theta_{(s,b)L}$ are the left-handed $(s,b)-\chi'_a$ mixing angles,
$\theta'_{(s,b)(L,R)}$ are the 
left-/right-handed $(s,b)-\chi''_a$ mixing 
angles,
and 
where we may take $\Lambda \sim {\cal O}(2\gev)$.

We can expand these operators in the basis
\bea
\frac{\alpha_{em} G_F}{\sqrt{2} \pi} V_{tb} V^*_{ts} \sum_{i,\ell} 
C_i^{bs\ell \ell} {\cal O}_i^{bs\ell \ell} ,
\eea
where 
\bea
{\cal O}_9^{bs\ell \ell} &=& (\bar s \gamma^\mu P_L b) (\bar \ell \gamma_\mu \ell) , 
\nonumber\\
{\cal O}_{10}^{bs\ell \ell} &=& (\bar s \gamma^\mu P_L b) (\bar \ell \gamma_\mu \gamma^5 \ell) , 
\nonumber\\
{\cal O}_9^{'bs\ell \ell} &=& (\bar s \gamma^\mu P_R b) (\bar \ell \gamma_\mu \ell) , 
\nonumber\\
{\cal O}_{10}^{'bs\ell \ell} &=& (\bar s \gamma^\mu P_R b) (\bar \ell \gamma_\mu \gamma^5 \ell) ,
\nonumber\\
{\cal O}_{S}^{bs\ell \ell} &=& m_b (\bar s P_R b) 
(\bar \ell \ell) ,
\nonumber\\
{\cal O}_{S}^{'bs\ell \ell} &=& m_b (\bar s P_L b) 
(\bar \ell \ell) ,
\nonumber\\
{\cal O}_{P}^{bs\ell \ell} &=& m_b (\bar s P_R b) 
(\bar \ell \gamma^5 \ell) ,
\nonumber\\
{\cal O}_{P}^{'bs\ell \ell} &=& m_b (\bar s P_L b) 
(\bar \ell \gamma^5 \ell).
\eea
Defining $C_i^U = C_i^{bsee}$ and 
$C_i^{NU} = C_i^{bs\mu\mu} - C_i^U$, we find 
\bea
\Delta C_{9}^{U} &=& (-146) (\sin \theta_{sL} \sin \theta_{bL} + \sin \theta'_{sL} 
\sin \theta'_{bL}) ,
\nonumber\\
\Delta C_{10}^{U} &=& (1.8 \times 10^3) 
(\sin \theta_{sL} \sin \theta_{bL} + \sin \theta'_{sL} 
\sin \theta'_{bL}) ,
\nonumber\\
\Delta C_9^{NU} &=& \Delta C_{10}^{NU} = 
(1.9 \times 10^8)
\sin \theta'_{sL} \sin \theta'_{bL} 
\left( \frac{m_{A'}}{\sqrt{2} V} \right)^2 ,
\nonumber\\
\Delta C_9^{'NU} &=&  
\Delta C_{10}^{'NU} =
(1.9 \times 10^8)
\sin \theta'_{sR} \sin \theta'_{bR} 
\left( \frac{ m_{A'}}{\sqrt{2} V} \right)^2 ,
\nonumber\\
\Delta C_{P}^{NU} &=&  
- \Delta C_{P}^{'NU} = -(2.0 \times 10^5 \gev^{-1})
\left(\frac{V}{10~\gev} \right)^{-2}
\nonumber\\
&\,& \times
\left(\sin \theta'_{sR} \sin \theta'_{bR} 
- (m_s/m_b) \sin \theta'_{sL} \sin \theta'_{bL} \right) , 
\nonumber\\
\Delta C_{S}^{NU} &=& (2.7\times 10^7 \gev^{-1})
\lambda''_b \sin \theta'_{sL} 
\frac{m_\mu}{m_b} \left(\frac{V}{10~\gev} \right)^{-1}, 
\nonumber\\
\Delta C_{S}^{'NU} &=& (2.7\times 10^7 \gev^{-1})
\lambda''_s \sin \theta'_{bL} 
\frac{m_\mu}{m_b} \left(\frac{V}{10~\gev} \right)^{-1} .
\nonumber\\
\eea
Since $\sin^2 \theta_W \sim 0.23$, the universal lepton vector coupling 
is negligible.  

We see that this scenario allows for several operators 
which contribute $b \rightarrow s \ell^+ \ell^-$ 
processes, with coefficients controlled by 
independently-tunable couplings and mixing angles.  
We find that we have freedom in the quark couplings, 
although the vector couplings to muons are only 
right-handed.

\subsection{Benchmark Scenarios}

We now use the allowed parameter space of $m_{A'}$ and $m_{\phi'}$ masses, as shown in Fig.~\ref{fig:VisibleSensitivity}, to explain the recently observed anomalies.    To study the implications of this scenario for flavor anomalies, we restrict our analysis to theoretically clean observables~\cite{Geng:2021nhg}, $R_K$, $R_{K^*}$, and $Br(B_s\rightarrow \mu^+\mu^-)$. $R_K$ and $R_{K^{*}}$ are defined as 
\begin{align}
\begin{split}
R_K &\equiv \frac{Br(B\to K \mu^+\mu^-)}{Br(B\to K e^+e^-)} ~,
\\
R_{K^*} &\equiv \frac{Br(B\to K^* \mu^+\mu^-)}{Br(B\to K^* e^+e^-)} ~.
\end{split}
\end{align}
Because of lepton flavor universality, the SM predictions for $R_K$ and $R_{K^{*}}$ are close to unity~\cite{Hiller:2003js,Bouchard:2013mia}, while the measurements have been consistently below the SM prediction~\cite{RKstar,RKstarBelle, Aaij:2021vac,Aaij:2014ora,Aaij:2019wad}. Recently, the LHCb Collaboration reported the most precise measurement of $R_K$ in the $q^2$ bin of 1.1 to 6 GeV$^2$ using the full Run-1 and Run-2 data sets shown in Eq.\ref{rklhcb}~\cite{Aaij:2021vac},
which deviates form the SM prediction by 3.1$\sigma$. The $R_{K^{*}}$ measurements~\cite{RKstar,RKstarBelle}
\begin{eqnarray}
R_{K^*} = 
\begin{cases}
0.660^{+0.11}_{-0.07}\pm 0.03  \,\, (2m_\mu)^2 < q^2 < 1.1~\mbox{GeV}^2 ~,
\\
0.685^{+0.11}_{-0.07}\pm 0.05 \,\, 1.1~\mbox{GeV}^2  < q^2 < 6~\mbox{GeV}^2 ~,
\end{cases}
\end{eqnarray}
disagree with the SM expectations at the $2.4\sigma$ and $2.5\sigma$ levels, respectively. In this study, we restrict ourselves to the central bin of $R_{K^{*}}$ measurement. It is known that explaining both bins with effective operators is extremely challenging, and we will wait for more data to confirm the energy dependency~\cite{Datta:2017ezo,Altmannshofer:2017bsz}. 
Together with other processes mediated by $b\rightarrow s \, \ell^+ \, \ell^-$ transitions, the tension is at least at the level of 4$\sigma$~\cite{Altmannshofer:2021qrr,Geng:2021nhg,Carvunis:2021jga}. LHCb also reported the measurement of the branching fraction of $B_s\rightarrow \mu^+ \mu^-$ using the full data set~\cite{LHCbseminar},

\begin{equation}
    Br(B_s\rightarrow \mu^+ \mu^-) = 3.09^{+0.46}_{-0.43}\textrm{(stat)}^{+0.15}_{-0.11}\textrm{(sysm)} 
    \times 10^{-9}.
\end{equation}
Together with the recent measurement by ATLAS~\cite{Aaboud:2018mst} and CMS~\cite{Sirunyan:2019xdu}, a decay rate smaller than the SM prediction is favored~\cite{Altmannshofer:2021qrr,Geng:2021nhg}.

In general, it is difficult to explain $g_{\mu}-2$, $R_{K^{(*)}}$, and $B_s\rightarrow \mu^+ \mu^-$ simultaneously with a vector mediator while respecting all current experimental constraints. The region that is consistent with $g_\mu-2$ is strongly constrained by beam dump and fixed target experiments for models such as $U(1)_{B-L}$~\cite{Bauer:2018onh}. In models such as $U(1)_{L_{\mu}-L_{\tau}}$~\cite{He:1990pn,He:1991qd}, a mediator around $10-100\mev$ with a coupling $g_{\mu\tau} \sim 
{\cal O}(10^{-4})$ - ${\cal O} (10^{-3})$ can accommodate the $g_\mu-2$ results. Heavier mediators require larger muon couplings and are heavily constrained by neutrino trident production at CCFR~\cite{Bauer:2018onh}. Then to accommodate the result of $R_K$ and $R_{K^{*}}$, a $bs$ coupling around ${\cal O} (10^{-10})$ - ${\cal O}(10^{-9})$ is required. In this scenario, a light mediator decays dominantly to neutrinos, and it contributes to the $B\rightarrow K^{*}X, X\rightarrow \nu\nu$ process. The couplings required to explain $R_{K^{(*)}}$ lead to $Br(B\rightarrow K^{*} X) \times Br(X\rightarrow \nu \nu)$ at least ${\cal O} (10^{-4})$. 
The measurement at Belle sets an upper limit on $Br(B\rightarrow K^{*}\nu\nu)$ of $5.5\times10^{-5}$ at 90$\%$ confident level~\cite{Lutz:2013ftz}, and thus exclude this simple scenario.  

The advantage of models with only right-handed lepton coupling such as $U(1)_{T3R}$ is that, due to the lack of left-handed neutrino couplings, the major experimental constraints, including CCFR and $B\rightarrow K^{*} \nu\nu$, do not apply to such scenarios.\footnote{In this scenario, there is a contribution to the $B\rightarrow K^{(*)}\nu\nu$ process from $B\rightarrow K^{(*)} A^{\prime}$, and $A^{\prime}\rightarrow \nu\nu$. These processes have hadronic form factor uncertainties~\cite{PhysRevD.71.014029,PhysRevD.95.094023,Khodjamirian:2010vf}. In addition, in the Belle and BaBaR analysis~\cite{Lutz:2013ftz,Lees:2013kla}, the invariant mass of the two neutrinos, $m_{\nu\nu}$ is required to be larger than about 2.5 GeV. Therefore, such measurements do not apply to the parameter space if $A^{\prime}$ dominantly decays into missing energy, i.e., $\nu_s\nu_s$. There would be constraints from the COHERENT and Crystal Barrel experiments for such a final state in this model for $m_{A^\prime}>30$ MeV. However for $m_{A^\prime}<30$ MeV all constraints are satisfied.  
}
But on the other hand, 
$U(1)_{T3R}$ models necessarily impose 
$C_9^{(\prime)NU} = C_{10}^{(\prime)NU}$, and this 
constraint makes it difficult 
to explain the $R_K$ and $R_{K^{*}}$, and $Br(B_S\rightarrow \mu^+\mu^-)$ measurements simultaneously. 
The $R_K$ and $R_{K^{*}}$ measurements prefers a negative $C_9^{bs\mu\mu}$, or a positive $C_{10}^{bs\mu\mu}$, and the smaller decay rate of $B_s \rightarrow \mu^+\mu^-$ favors a positive $C_{10}^{bs\mu\mu}$, or a negative $C_{10}^{\prime bs\mu\mu}$. To explain $R_K$ and $R_{K^{*}}$ with a  positive $C_{10}^{bs\mu\mu}$, which is favored by $B_s\rightarrow \mu^+\mu^-$, implies a negative $C_{9}^{bs\mu\mu}$. 
Since $C_9^{NU} = C_{10}^{NU}$, 
a negative $C_{9}^{bs\mu\mu}$ and a positive $C_{10}^{bs\mu\mu}$  imply a negative non-universal part and a positive universal part. Then a positive $C_{10}^{bsee}$ will leave the $R_K$ and $R_K^{*}$ unexplained. 

Therefore, we consider the following two additional 
scenarios. In the first scenario, we introduce scalar and pseudo-scalar couplings. We rely on the scalar and pseudo-scalar operators to explain $Br(B_S\rightarrow \mu^+\mu^-)$, while $R_{K^{(*)}}$ can be fixed by other operators. In the second scenario, we include the prime operators, which only contain the non-universal part, so that the contributions are generated from both left-handed and right-handed quark couplings. 

In Table~\ref{table:BM}, we present four benchmarks. For all benchmark points, we calculate the corresponding flavor observables with flavio~\cite{Straub:2018kue}, and also calculate the SM pull, defined as $\sqrt{\Delta\chi^2}$, using the clean observables only, to show how well those three measurements can be described and how significant the deviation is from the SM. When we calculate the SM pull, we only include the LHCb results for simplicity. The Belle 
measurements of $R_{K^{(*)}}$ have significantly larger 
uncertainties compared to the LHCb results~\cite{RKstar,RKstarBelle,Aaij:2021vac}, while the $B_s\rightarrow \mu^+\mu^-$ measurements by ATLAS and CMS correlates with $B_d\rightarrow\mu^+\mu^-$~\cite{Aaboud:2018mst,Sirunyan:2019xdu}. 
The energy-dependant behavior in $R_{K^*}$ is beyond the scope of this study, so we only list the value of $R_{K^*}$ in the central $q^2$ bin, as indicated by numbers in the bracket. Here $q^2$ is defined as the invariant mass-squared of the dimuon system.

The first three benchmarks correspond to the first scenario in which scalar and pseudo-scalar operators are responsible for $Br(B_S\rightarrow \mu^+\mu^-)$. In BMA, we include the scalar operators, and in BMB, and in BMC, we include both scalar and pseudo-scalar operators. For BMA, $R_K$ and $B_s\rightarrow\mu^+\mu^-$ agree with the LHCb results within 1$\sigma$, and $R_{K^{*}}$ agree with the LHCb results within 2$\sigma$, and the SM pull is 4.4$\sigma$ for BMA. For the second benchmark BMB, all three observables agree with the LHCb measurements within 1$\sigma$, with a SM pull of 4.6$\sigma$. For BMC, $R_K$ and $B_s\rightarrow\mu^+\mu^-$ agree with the LHCb results within 1$\sigma$, while $R_{K^{*}}$ is SM like. 

BMD corresponds to the second scenario. Introducing only left-handed quark couplings does not provide a good explanation for all three measurements. As discussed above, a pure left-handed quark coupling will leave either $R_{K^{*}}$ or $Br(B_s\rightarrow\mu^+\mu^-)$ unexplained. 
Therefore, we further include the non-universal, primed operators. $R_K$ and $B_s\rightarrow\mu^+\mu^-$ agree with the LHCb results within 1$\sigma$, and $R_{K^{*}}$ agree with the LHCb results within 2$\sigma$, and the SM pull is 4.7$\sigma$ for BMD. 

In BMA and BMC, with a negative $C_9$ and a positive $C_{10}$, both electron and muon modes are suppressed compared to the SM, and $R_K$ and $R_{K^{*}}$ are explained by suppressing the muon mode even more from the non-universal part. In BMB and BMD, both electron and muon modes are enhanced compared to the SM, and $R_K$ and $R_{K^{*}}$ are explained by increasing the muon mode less from the non-universal part. For all benchmark scenarios we considered, the $Z$-mediated contributions to $B\rightarrow K^{(*)} \nu\nu$, and $B\rightarrow K^{(*)} ee$ are well below the current upper limit~\cite{Lutz:2013ftz,Grygier:2017tzo,Aaij:2013hha}, and contributions to $B_s-\bar{B_s}$ mixing are negligible as $bsbs$ operators are very small in the region of interest.

\begin{table*}[ht]
\captionsetup{justification   = RaggedRight,
             labelfont = bf}
             \caption{Details of the four benchmark points 
             described in the text.  The first five rows 
             present the values of the coefficients 
             $C_{10}^{U}$, $C_{9,10}^{NU}$, 
             $|C_s - C'_s|$ (in units of $\gev^{-1}$), 
            $|C_p - C'_p|$ (in units of $\gev^{-1}$), 
            and $C^{'NU}_{9,10}$.  Rows 6-8 present 
            predictions for $R_K$, 
            $R_{K^*}$ (in the $q^2 \in [1.1,6] \gev^2$ 
            bin), and $Br(B_s \rightarrow \mu^+\mu^-)$.  Row 
            9 presents the SM pull of each benchmark point.
             \label{table:BM}}

\begin{ruledtabular}
\begin{tabular}{c|c|c|c|c}
&BMA&BMB &BMC&BMD \\
\hline
$C_{10}^{U}$ &4.85& -5.86&2.7 &-5.67\\
$C_{9,10}^{NU}$&-0.30 &3.65 &-0.8&4.55\\
$|C_s-C_s^{\prime}|$ GeV$^{-1}$& 0.033& 0.024 & 0.011& - \\
$|C_p-C_p^{\prime}|$ GeV$^{-1}$& - &0.030 &0.043 & - \\
$C_{9,10}^{\prime NU}$& -  & - &- &-1.28\\
\hline
$R_K$& 0.82 & 0.87&0.86 &0.87\\
$R_K^{*}[1.1,6]$ & 0.83&0.78 &0.97 &0.89 \\
$Br(B_s\rightarrow \mu^+\mu^-)$&3.36$\times 10^{-9} $ &3.05$\times 10^{-9}$ & 2.67$\times10^{-9}$&3.34$\times 10^{-9}$ \\
SM pull&4.4$\sigma$ &4.6$\sigma$ &3.8$\sigma$ & 4.2$\sigma$ \\

 \end{tabular}
\end{ruledtabular}
             
\end{table*}

\begin{table*}[ht]
\captionsetup{justification   = RaggedRight,
             labelfont = bf}
             \caption{Predictions for observables for the four 
benchmark points described in the text (columns 4-7),
along with the 
Standard Model prediction (3rd column) and the 
measured value with uncertainties (2nd column). The uncertainties, from left to right, are statistical, systematic and due to the normalisation mode~(for the last two only).  Rows 1-3 consider 
$Br (B^+ \rightarrow K^{*+}\mu^+\mu^-) (q^2 \in [15,19]\gev^2)$, $Br (B^0 \rightarrow K^{0}\mu^+\mu^-) (q^2 \in [15,19]\gev^2)$, 
and
$Br (B^+ \rightarrow K^{+}\mu^+\mu^-) (q^2 \in [15,22]\gev^2)$, 
respectively, all in units of $10^{-8}$.  Row 4 considers 
$dBr(B_S \rightarrow \phi \mu^+\mu^-) / dq^2$, in units of 
$10^{-8} \gev^{-2}$, averaged over 
$q^2 \in [1,6] \gev^2$, while row 5 considers  
$dBr(\Lambda_b^0 \rightarrow \Lambda \mu^+\mu^-) / dq^2$, in units of 
$10^{-7} \gev^{-2}$, averaged over 
$q^2 \in [15,20] \gev^2$.
\label{table:predictions}}

\begin{ruledtabular}
\begin{tabular}{c|c|c|c|c|c|c}
Observable&Measured Value&SM&BMA& BMB &BMC&BMD \\
\hline
$Br(B^{+}\rightarrow K^{*+}\mu^+\mu^-)$($10^{-8}$)[15.0,19.0] &15.8$^{+3.2}_{-2.9}\pm 1.1$ ~\cite{Aaij:2014pli} & 26.8$\pm$3.6 &  7.80&82.9&10.4& 92.4\\
$Br(B^{0}\rightarrow K^{0}\mu^+\mu^-)$ ($10^{-8}$)[15.0,22.0]&6.7$\pm$1.1$\pm$0.4~\cite{Aaij:2014pli}&9.8$\pm$1.0&3.31&30.4&4.15&29.4\\ 
$Br(B^{+}\rightarrow K^{+}\mu^+\mu^-)$($10^{-8}$) [15.0,22.0] &$8.5\pm0.3\pm0.4$~\cite{Aaij:2014pli}&$10.7\pm1.2$&3.59&33.0&4.5&32.0\\
$\frac{dB(B_S\rightarrow\phi \mu^+\mu^-)}{dq^2}$ ( $10^{-8}$ GeV$^{-2}$)[1.0,6.0]&2.57$^{+0.33}_{-0.31}\pm0.08\pm0.19$~\cite{Aaij:2015esa}&$4.81\pm0.56$&1.60&16.8&2.28&18.7\\
$\frac{dB(\Lambda_b^0\rightarrow\Lambda\mu^+\mu^-)}{dq^2}$ ($10^{-7}$ GeV$^{-2}$) [15,20] &$1.18^{+0.09}_{-0.08}\pm0.03\pm0.27$~\cite{Aaij:2015xza}&$0.71\pm0.08$& 2.19&2.28 &0.29&2.48\\ 
 \end{tabular}
\end{ruledtabular}
\end{table*}

We have  listed in Table~\ref{table:predictions} the predictions of this model for other observables with large 
theoretical uncertainties. We also list the current experimental value and the SM predictions, calculated by flavio~\cite{Straub:2018kue} for references. Currently, those observables are measured with 3 fb$^{-1}$ of data. The numbers in the bracket show the range of the the invariant mass-squared of the dimuon system, $q^2$. The uncertainties in the experimental value, from left to right, are statistical, systematic and due to the normalisation mode~(for the last two only). As discussed above, in BMA and BMC, the muon modes are suppressed, as indicated by the current experiments, while in BMB and BMD, the muon modes are enhanced compared to the SM.

Because of the universal contribution, in order 
to accommodate the experimental value of $R_K$ and $R_{K^{*}}$, sizable deviations from SM predictions 
are expected for the unclean observables. Models with 
smaller Wilson coefficients, such as in BMB, lead to a good explanation to the unclean observables. These observables, however, involve form factor related uncertainities, which 
lead to large corrections to the model predictions. 
If these theoretical uncertainties can be brought under 
control, then model predictions can be more meaningfully 
compared to data.

Below the dimuon threshold, $A'$ may decay to $e^+ e^-$ via kinetic mixing with the photon. In processes such as $B\rightarrow K e^+e^-$ (which contain hadronic form factor uncertainties~\cite{PhysRevD.71.014029,PhysRevD.95.094023,Khodjamirian:2010vf}), an  $A'$ can be produced on-shell via  $B \rightarrow K A'$, with the  $A'$ decaying to an $e^+e^-$ pair, potentially leading to a signal in a resonance search. Although LHCb does have constraints on the  dark photon using $\ell \ell$ resonance searches, it has no constraints on the $e^+e^-$ decay mode in the energy range of interest. LHCb constraints use the $\mu^+ \mu^-$ final state for $m_{A'}$ $\geq 2m_\mu$~\cite{LHCb:2020ysn}. For $B \rightarrow K^\ast \ell \ell$ modes~\cite{LHCb:2019hip,LHCb:2020dof}, LHCb performs a resonance ($e^+e^-$) analysis only  for $q^2>6$ GeV$^2$, using the $J/\psi\rightarrow e^+e^-$ channel. Below 6 GeV$^2$, there exists no resonance study providing the distribution 
$m_{\ell \ell} (q^2)$.  The  minimum  angular separation between $e^+$ and $e^-$ is also not given ($e^+e^-$ is quite collimated for such a low $A'$ mass, as in our scenario). LHCb also performs non-resonance studies of the invariant masses $m(K\pi \ell \ell)$ and $m(K\ell \ell)$ for the $B\rightarrow K^{\ast} \ell \ell$ and $B \rightarrow K{\ell \ell}$ decay modes, respectively which does not constrain our model.
Since we consider $m_{A'} \sim 100\mev$,  one needs a dedicated resonance study with the $e^+e^-$ final state to obtain constraints.  Currently we do not have any constraint from LHCb  on this resonance channel. 


In this setup, we introduced mixing in the second and third generation down-type quark sector via heavy quarks, and we have discussed the associated  predictions for flavor-changing neutral currents.  But this scenario does not generate contributions 
to the CKM matrix.  To do so, we would need to turn on mixing among all the generations of up- 
and down-type quarks~\cite{Berezhiani:1983hm,Chang:1986bp,Davidson:1987mh,DePace:1987iu,Rajpoot:1987fca,Babu:1988mw,Babu:1989rb,Babu:2018vrl}.


 \section{Conclusion}
\label{sec:Conclusion}

Scenarios in which first-/second-generation 
right-handed SM fermions are charged under 
$U(1)_{T3R}$ are particularly interesting.  Among 
all scenarios involving new gauge groups, this 
scenario is distinctive because the coupling of 
the new particles to the SM is constrained from 
below; because the new symmetry protects fermion 
masses, the coupling of the symmetry-breaking field 
to SM fermions is proportional to the fermion mass.  
This yields an attractive scenario in which the 
symmetry-breaking naturally sets not only the light 
SM fermion masses, but also the mass scale of the 
dark sector, naturally pointing to sub-GeV dark 
matter.  But the other side of this coin is that the 
symmetry-breaking field necessarily has a large 
coupling to SM fields, as it is proportional to the
ratio of SM fermion mass and the symmetry-breaking 
scale, which is presumed to be not large.
This coupling is 
inherited by the dark Higgs and the Goldstone 
mode (which is absorbed into the dark photon 
longitudinal polarization).  This scenario thus 
faces tight constraints from searches for these 
mediators, and only a narrow range of parameter 
space is still viable.

These couplings are particularly relevant to the 
corrections to $g_\mu-2$, as both the dark Higgs 
and dark photon yield corrections which are 
roughly two orders of magnitude too large.  But 
within the small region of parameter space which 
is allowed by other experiments, the corrections 
from the dark photon and the dark Higgs can cancel, 
yielding an overall contribution which matches the 
latest measurements from Fermilab.

This scenario necessarily leads to lepton flavor 
non-universality arising from low-energy physics. 
Moreover, UV completions of this scenario can easily 
accommodate quark flavor-violation.  These are the 
required ingredients for explaining  the anomalies 
in $R_{K^{(\ast)}}$ observation. 
We show that we can have necessary operators to explain the anomalies after satisfying $B_s\rightarrow\mu^+\mu^-$ constraint in the allowed parameter space  where the $g_\mu -2$ 
anomaly is also explained. In general, it is not easy to explain both anomalies after satisfying various constraints. Various neutrino related measurements restrict the parameter space of the models which utilize left handed muons to solve the $R_{K^{(\ast)}}$ puzzle. However, this problem is ameliorated in the context of  the $U(1)_{T3R}$ model due to the absence of the left-handed neutrino couplings of $A'$. We also list predictions for a few more observables which  can be tested in the future. The future measurements of $R_{K^{(\ast)}}$ would be crucial to probe this scenario. In addition, as an example, we show a possible UV completion of this scenario based on the universal seesaw mechanism. The new heavy vector-like fermions introduced can lead to strong first-order electroweak phase transitions and the corresponding gravitational wave signal provides an additional probe to this scenario~\cite{Angelescu:2018dkk}. As a future work, the cosmological dynamics behind this scenario will be studied in further detail.

It is interesting to probe the allowed parameter space of this model with future 
experiments.  Future searches at experiments such 
as FASER, SeaQuest and SHiP may find evidence for 
the displaced decays of $A' \rightarrow e^+ e^-$.  But 
the difficulty is that, the very fact that the dark 
photon and dark Higgs contributions to 
$g_\mu -2$ must be canceled 
against each other shows that they were both large, 
leading to an $A'$ decay rate which is larger than 
usually expected.  As a result, the $A'$ often decays 
before it reaches a displaced detector.  To test this 
scenario definitively, it would be best to have an 
experiment with a shorter distance from the target to 
the displaced detector.

\begin{figure}[h]
\centering
\includegraphics[height=5.5cm,width=8cm]{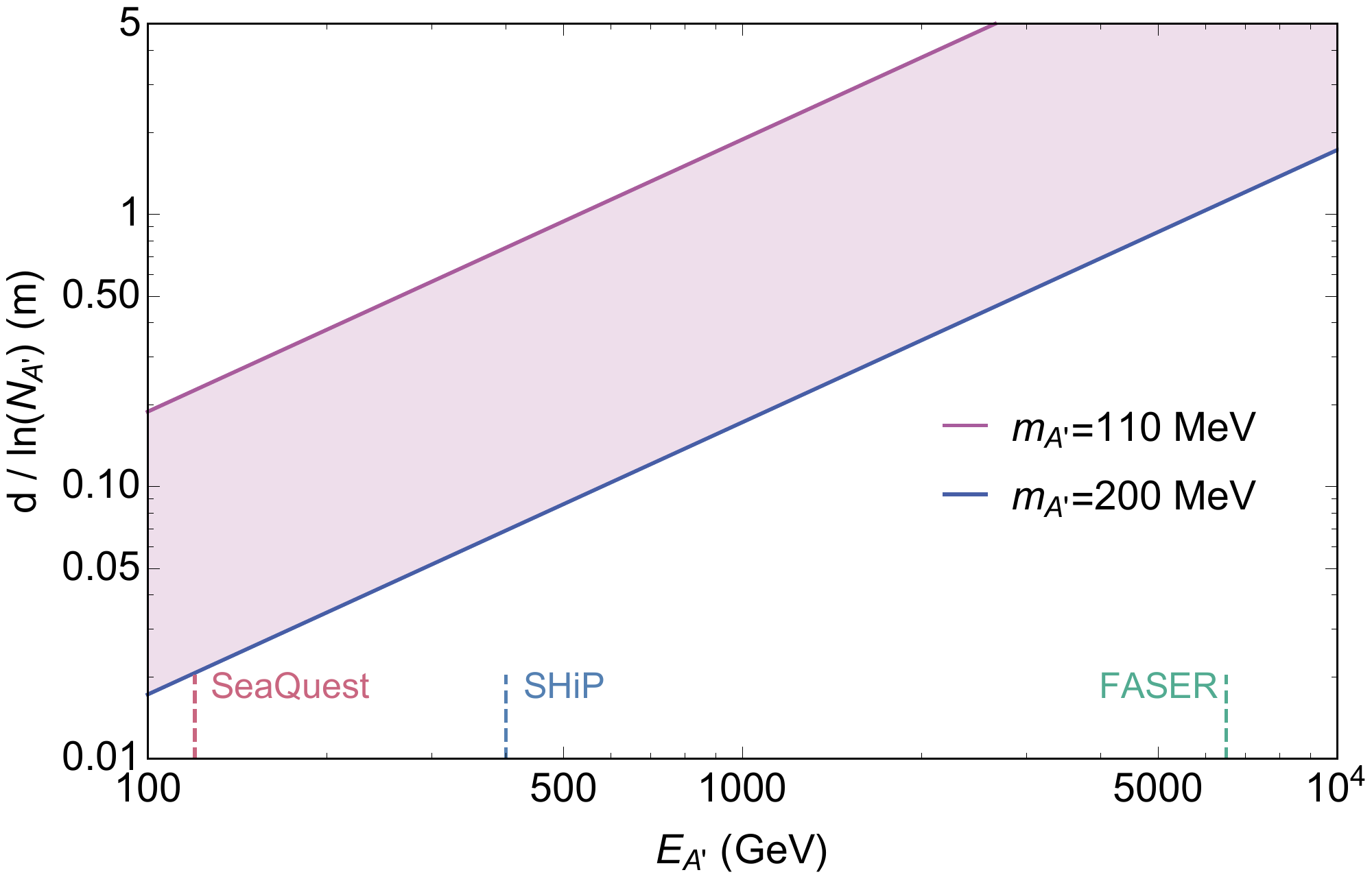}
\captionsetup{justification   = RaggedRight,
             labelfont = bf}
\caption{\label{fig:dvsEAplot} A rough estimate of 
maximum $d/\ln(N_{A'})$ necessary for an experiment to 
be able to probe this scenario for 
$m_{A'} \in [110\mev, 200\mev]$, as a function of the 
maximum $A'$ energy produced by the experiment.  
$d$ is the displacement of the detector from the 
beam dump, and $N_{A'}$ is the number of $A'$ at 
energy $E_{A'}$ produced 
in a beam aimed at the detector.  The maximum 
$A'$ energies 
of FASER, SHiP and SeaQuest are also shown. }
\end{figure}

We can consider the properties needed by a future 
displaced detector experiment to probe these models.  
If $N_{A'}$ is the number of $A'$ at characteristic energy $E$ produced by the 
beam which would reach the detector if $A'$ were 
stable, the number which reach the a detector 
a distance $d$ away is $N_{A'} \exp [- d / d_{dec}]$, 
where $d_{dec}$ is the decay length for an $A'$ of 
energy $E_{A'}$.  If $d_{decay} \ll d$, then most 
most $A'$ which reach the detector will decay 
shortly after.  So if we set this number to be of 
order unity, as a rough estimate of the number of 
$A'$ reaching the detector necessary for a signal 
to be detected above negligible background, then 
we find $d_{dec} =d / \ln (N_{A'})$.  $d_{dec}$ is 
determined by the model, but $d / \ln (N_{A'})$ 
is entirely determined by the properties of 
the instrument, and is a function of the maximum 
typically energy of the produced $A'$.  We plot this 
quantity as a function of $E_{A'}$ in 
Figure~\ref{fig:dvsEAplot}.

An alternative would be to search for visible decays 
of the $\phi'$ ($\phi' \rightarrow \gamma \gamma$).  
Searches for this decay channel require detailed 
study of $\phi'$ production mechanisms.  It would 
be interesting to perform a more detailed study of 
the sensitivity of displaced decay experiments.  
Alternatively, one could search for the 
central production of $\phi'$ at the LHC; where 
it could appear either as missing energy, or as a 
monophoton or diphoton signal.  
It maybe possible to search for these signal in events where $\phi'$  
receives a large transverse boost against a recoiling 
photon or jet.  It would be interesting to study this 
possibility in greater detail.

\begin{acknowledgments}
We would like to thank L.~Darme and W.~Parker for helpful discussions. The work of BD and SG are supported in part by the DOE Grant No.~DE-SC0010813.
The work of JK is supported in part by DOE Grant No.~DE-SC0010504. The work of PH is supported by University of Nebraska-Lincoln, and National Science Foundation under grant number PHY-1820891. We have used the package TikZ-Feynman~\cite{Ellis:2016jkw} to generate the Feynman diagram of Fig.~\ref{fig:g2T3Rscalar}, and \ref{fig:RK}.
\end{acknowledgments}

\appendix

\section{Model Description}

The gauge symmetry of our model is $SU(3)_C\times SU(2)_L \times U(1)_Y \times U(1)_{T3R}$. The electric charge is defined as $Q = T3L + Y$, such that the new gauge group $U(1)_{T3R}$ is not connected to electric charge.

In addition to the light fields $\phi$, $A'$,
$\eta$ and $\nu_R$ (discussed in detail in the 
text and Ref.~\cite{Dutta:2019fxn}), we add a set of heavy 
fermions $\chi_{u,d,\mu,\nu}$ which are singlets 
under $SU(2)_L$ and $U(1)_{T3R}$, and have same 
quantum numbers under $SU(3)_C$ and $U(1)_Y$ as 
$u$, $d$, $\mu$ and $\nu$, respectively.  These fermions 
will mix with the fermions charged under 
$U(1)_{T3R}$, generating the mass terms 
and couplings of the light fermions through a 
high-scale seesaw mechanism.
The charge assignment of relevant 
particles are given in Table.~\ref{table:charge}. 

\begin{table}[h]
 \captionsetup{justification   = RaggedRight,
             labelfont = bf}
 \caption{ \label{table:charge} The 
charges of the fields 
under the gauge groups of the model. For the fermionic fields, the shown charges are for the left-handed component of each Weyl spinor.  }
\begin{ruledtabular}
\begin{tabular}{ ll }

Particle  & $SU(3)_C\times SU(2)_L \times U(1)_Y \times U(1)_{T3R}$   \\\hline
$\chi_{uL}$ & $(3,1, 2/3,0)$\\
$\chi_{d L}$ & $(3,1, -1/3,0)$\\
$\chi_{\mu L}$ & $(1,1,-1,0)$\\
$\chi_{\nu L}$ & $(1,1,0,0)$\\
$\chi_{uR}^c$ & $(3,1, -2/3,0)$\\
$\chi_{d R}^c$ & $(3,1, 1/3,0)$\\
$\chi_{\mu R}^c$ & $(1,1,1,0)$\\
$\chi_{\nu R}^c$ & $(1,1,0,0)$\\
$q_L$ & $(3,2,1/6,0)$\\
$u_R^c$ & $(3,1,-2/3,-2)$\\
$d_R^c$ & $(3,1,1/3,2)$\\
$l_L$ & $(1,2,-1/2,0)$\\
$\mu_R^c$ & $(1,1,1,2)$\\
$\nu_R^c$ & $(1,1,0,-2)$\\
$\eta_L$ & $(1,1,0,1)$\\
$\eta_R^c$ & $(1,1,0,-1)$\\
$H$ & $(1,2,1/2,0)$\\
$\phi$ & $(1,1,0,2)$\\

\end{tabular}
\end{ruledtabular}
\end{table}

The scalar potential can be written as 
\bea  
V &=& m_{H}^2 H^\dagger H + m_{\phi}^2 \phi^* \phi  + \lambda_H (H^\dagger H)^2 + \lambda_\phi (\phi^* \phi)^2 \nonumber\\ & &+ \lambda (H^\dagger H)(\phi^* \phi).  
\eea

Both scalar fields will get vevs, $\langle 
H \rangle = v/\sqrt{2}$ and $\langle \phi \rangle = V$. After the spontaneous symmetry breaking, the scalar fields can be written as, 
\bea &H & =   \left( \begin{array}{c}  {G}^+  \\  \frac{1}{\sqrt{2}}(v +\rho_0+iG_0)  \end{array} \right) 
\nonumber  \\
&\phi& = V+\frac{1}{\sqrt{2}} (\rho_\phi+i G_{\phi 0 })~.~\, 
\eea
 
 There are total 6 scalar degrees of freedom (dof), out of which 4 are absorbed into the 
 longitudinal polarizations of the 
 $W^{\pm}, Z$ and $A^\prime$ gauge bosons. The remaining 2 are the physical Higgs and dark Higgs scalars. The CP-even states $\rho_0$ and $\rho_\phi$ mix and give rise to the two physical neutral scalar $h$ and $\phi^\prime$ with masses $m_h$ and $m_\phi^\prime$ respectively. We identify $h$ as the SM Higgs boson. The two  physical  neutral scalars in terms of the interaction states are given as , \begin{equation} \left( \begin{array}{c} h \\ \phi^\prime \end{array} \right) =  \left( \begin{array}{cc} \cos \alpha & -\sin  \alpha \\ \sin  \alpha &\cos  \alpha \end{array} \right) \left( \begin{array}{c} \rho_0 \\ \rho_\phi \end{array} \right) ~,~\,\end{equation} where $\alpha$ is the mixing angle.

The decay rate for $h \rightarrow \phi' \phi'$ 
is constrained by LHC data.  To remain consistent 
with this data, one must assume that $\lambda$ 
(equivalently, $\alpha$) is small.

The renormalizable Yukawa sector Lagrangian of the UV-complete model in the interaction basis is given by,
\bea -\mathcal{L}_{\text{Y}} &=& \lambda_{Lu} \bar{q}_L^\prime \chi_{uR}^\prime \tilde{H} + \lambda_{Ld} \bar{q}_L^\prime \chi_{dR}^\prime H   \nonumber\\ &+& \lambda_{L\nu} \bar{l}_L^\prime \chi_{\nu R}^\prime \tilde{H} + \lambda_{Ll} \bar{l}_L^\prime \chi_{\mu R}^\prime H +\lambda_{Ru} \bar{\chi}_{uL}^\prime u_{R}^\prime \phi^*  \nonumber\\ &+&\lambda_{Rd} \bar{\chi}_{dL}^\prime d_{R}^\prime \phi  +\lambda_{
R \nu} \bar{\chi}_{\nu L}^\prime \nu_{R}^\prime \phi^* +\lambda_{Rl} \bar{\chi}_{\mu L}^\prime \mu_{R}^\prime \phi \nonumber \\ &+& m_{\chi_u} \bar{\chi}_{uL}^\prime \chi_{uR} + m_{\chi_d} \bar{\chi}_{dL}^\prime \chi_{dR} +m_{\chi_\nu} \bar{\chi}_{\nu L}^\prime \chi_{\nu R} \nonumber \\ &+& m_{\chi_\mu} \bar{\chi}_{\mu L}^\prime \chi_{\mu R} + m_D \bar{\eta}_R \eta_L + \frac{1}{2} \lambda_{\eta L} \bar{\eta}^c_L \eta_L \phi  \nonumber \\ &+& \frac{1}{2} \lambda_{\eta R} \bar{\eta}^c_R \eta_R \phi^* + H.c.~,~\, \eea

The fermionic flavor eigenstates will mix and give rise to the mass eigenstates. The mass matrix in the flavor eigenstate basis is given by, 
\bea 
M_f = \left( \begin{array}{cc} 0 & \frac{\lambda_{Lf} v}{\sqrt{2}} \\ \lambda_{Rf} V & m_{\chi^\prime_f} \end{array} \right). 
\eea 
The diagonalization of the fermionic mass matrix using the seesaw mechanism gives two mass eigenstates. The lightest  mass eigenstates is the SM fermion while the heavier one is the physical vector-like fermion. The mass term for the SM fermion is, 
\bea 
m_f  = \frac{\lambda_{Lf} \lambda_{Rf} vV}{ \sqrt{2} m_{\chi^\prime_f}}, 
\eea 
and the physical vector-like fermion mass is 
\bea m_{\chi_f} \simeq m_{\chi^\prime_f}.  
\eea 
The neutrino mass matrix will be more complicated $3\times3$ matrix since they can also get Majorana maases as both $\nu^\prime_R$ and $\chi_\nu^\prime$ are uncharged under the unbroken SM gauge groups. The fermion mass eigenstates can be written in terms of the flavor eigenstates as follow, 
 \begin{equation} 
 \left( \begin{array}{c} f_{L,R} \\ \chi_{f_{L,R}} \end{array} \right) =  \left( \begin{array}{cc} \cos \theta_{f_{L,R}} & \sin  \theta_{f_{L,R}} \\ -\sin  \theta_{f_{L,R}} &\cos  \theta_{f_{L,R}} \end{array} \right) \left( \begin{array}{c} f^\prime_{L,R} \\ \chi_{f^\prime_{L,R}} \end{array} \right) ~,~\, 
 \end{equation}  
 where $\theta_{f_{L,R}}$ are the mixing angles. In the high-scale seesaw limit, $m_{\chi_f} \gg \lambda_{Lf} v/2 $ we get, 
 \bea \theta_{f_{L}} \simeq \tan^{-1}\left[  \frac{\lambda_{Lf} v} { \sqrt{2}m_{\chi_f} } \right] ,
 \eea and if $m_{\chi_f} \gg \lambda_{Rf} V $ then,
 \bea \theta_{f_{R}} \simeq \tan^{-1} \left[ \frac{\lambda_{Rf} V} { m_{\chi_f} } \right] .
 \eea
 
 The mass matrix of the $\eta$ field contains both Dirac terms, $m_D$, and Majorana terms, $m_M$. The Majorana term, $m_M$, is proportional to the vev $V$ as $m_M = \lambda_M V$, where we assume that $\lambda_M = \lambda_{\eta L}=\lambda_{\eta R}$. We further assume that $m_M \gg m_D$. We get two Majorana fermions, $\eta_1$ and $\eta_2$, with masses $m_1=m_M-m_D$ and $m_2=m_M+m_D$ respectively. 
 
 In the low-energy effective field theory defined below the electroweak symmetry breaking scale, the interactions of the SM fermions and the dark matter fields, $\eta$, with the $\phi^\prime$ is given by,
 \bea 
 -\mathcal{L}  
 &=& \frac{m_f}{\sqrt{2} V} \bar{f}f \phi^\prime +  \frac{m_1}{2 \sqrt{2} V} \bar{\eta}_1 \eta_1 \phi^\prime 
 \nonumber\\  
 &\,&+  \frac{m_2}{2 \sqrt{2} V} \bar{\eta}_2 \eta_2 \phi^\prime .
 \eea

\bibliographystyle{apsrev4-1.bst}
\bibliography{g2T3R.bib}

\end{document}